\documentclass[12pt]{article}
\usepackage{amsmath,amssymb,graphicx,mathrsfs,slashed}
\usepackage[usenames, dvipsnames]{color} 
\newcommand{\be}{\begin{equation}}
        \newcommand{\ee}{\end{equation}}
\newcommand{\bea}{\begin{eqnarray}}
        \newcommand{\eea}{\end{eqnarray}}

\def\({\left(} \def\){\right)}

\renewcommand{\baselinestretch}{1.5}
\begin{document}
\title{\vspace{-1.8in}
{The state of Hawking radiation is non-classical}}
\author{\large Ram Brustein${}^{(1)}$, A.J.M. Medved${}^{(2,3)}$, Yoav Zigdon${}^{(1)}$
\\
\vspace{-.5in} \hspace{-1.5in} \vbox{
\begin{flushleft}
$^{\textrm{\normalsize
(1)\ Department of Physics, Ben-Gurion University,
Beer-Sheva 84105, Israel}}$
$^{\textrm{\normalsize (2)\ Department of Physics \& Electronics, Rhodes University,
Grahamstown 6140, South Africa}}$
$^{\textrm{\normalsize (3)\ National Institute for Theoretical Physics (NITheP), Western Cape 7602,
South Africa}}$
\\ \small \hspace{1.07in}
ramyb@bgu.ac.il,\ j.medved@ru.ac.za,\ yoavzig@post.bgu.ac.il
\end{flushleft}
}}
\date{}
\maketitle
\begin{abstract}
We show that the state of the Hawking radiation emitted from a large Schwarzschild black hole (BH) deviates significantly from a classical state, in spite of its apparent thermal nature. For this state, the occupation numbers of single modes of massless asymptotic fields, such as photons, gravitons and possibly neutrinos, are small and, as a result, their relative fluctuations are large. The occupation numbers of massive fields are much smaller and suppressed beyond even the expected Boltzmann suppression. It follows that this type of thermal state cannot be viewed as classical or even semiclassical. We substantiate this claim by showing that, in a state with low occupation numbers, physical observables have large quantum fluctuations and, as such, cannot be faithfully described by a mean-field or by a WKB-like semiclassical state. Since the evolution of the BH is unitary, our results imply that the state of the BH interior must also be non-classical when described in terms of the asymptotic fields. We show that such a non-classical interior cannot be described in terms of a semiclassical geometry, even though the average curvature is sub-Planckian.
\end{abstract}
\newpage
\renewcommand{\baselinestretch}{1.5}\normalsize
\tableofcontents
\section{Introduction}

The main objective of this paper is to demonstrate the strongly non-classical nature of the Hawking radiation that is emitted from a large Schwarzschild black hole (BH). Similar claims about the non-classicality of Hawking radiation, in spite of its apparent thermal nature, have been put forth in an earlier article \cite{density}, and the possible consequences of this non-classicality
were subsequently discussed in \cite{inny,strungout,emerge}. In these
studies, however, no detailed evidence was provided except to point out that the occupation numbers of the Hawking modes are inevitably small. Yet, after surveying the literature, we were challenged to find a clear statement in support of our proposed relationship between low occupation numbers and non-classicality in a thermal state. (One notable exception can be found in Chapter 8 of \cite{duncan}.) The aim of this paper is to correct this omission.

The typical first step in a work of this nature would be to provide the reader with a precise mathematical definition of a non-classical state. It turns out, however, that there is no consensus viewpoint on such a definition, nor is there any ``one-size-fits-all'' diagnostic that can be used to distinguish between the non-classical and classical realms. Qualitatively, one can rely on the correspondence principle: When the state occupies a volume in phase space that is much larger than $\hbar$, then it can be said to ``behave classically". Some examples of this are states of quantum fields with large occupation numbers  and  highly excited states in quantum mechanics.

Quantitatively, we would like to know whether a state can be approximated by the WKB or mean-field approximations or some other similar type of semiclassical approximation. If such approximations are not valid, the state in question would fail to have a faithful semiclassical description, never mind a classical one. Formally, this amounts to identifying a dimensionless and effective ``$\hbar$'', the characteristic expansion parameter for the expectation values of some class of observables. As an expansion parameter, a dimensionless $\hbar$ is required to be less than unity; otherwise, the desired (semi-) classical approximation is bound to  fail.

Here, as a criterion for distinguishing between semiclassical and  non-classical states, we will be using the strength of the quantum fluctuations about the average values of certain observables. The dimensionless $\hbar$ will then  be a parameter that determines the relative strength of such  fluctuations. For example, in a one-dimensional WKB expansion, the dimensionless $\hbar$ is the ratio of the quantum de-Broglie wavelength $\lambda(x)= \frac{\hbar}{\sqrt{2m(E-V(x))}}$ to some geometric scale $\ell_{cl}$ which is determined by the potential $V$; that is, $\frac{\lambda}{\ell_{cl}}$. In a $\phi^4$ scalar field theory in four dimensions, it is rather the dimensionless coupling constant.

In gravity, the standard choice of  dimensionless $\hbar$ is $\;G_NE^2=E^2/M_P^2\;$, where $G_N$ is Newton's constant, $E$ is a typical energy scale and $M_P$ is the Planck mass \cite{Donaghue},\cite{Burgess}. As a result, there is  widely held impression that the effects  of quantum gravity  will only be significant when the energy is Planckian.  But this is not always the case, as we show later.

As will be highlighted in Section~2, the relevant dimensionless $\hbar$ for the state of the Hawking radiation is the inverse of the mode occupation number $1/\langle n_k\rangle$. Since these occupation numbers are small in comparison to unity, the dimensionless $\hbar$ is large, leading to the conclusion that the state of the Hawking radiation is non-classical. We will show in Section~3 how this non-classicality implies large geometrical fluctuations, even in regions of spacetime with a small average background curvature. It  will also pave the way to an interesting conclusion about the state of the BH interior. As will be explained in Section~4, the density matrices of the BH interior and the  Hawking radiation share a common set of non-vanishing eigenvalues, as either  of these systems acts as  the purifier of the other. One can then  infer that the interior is similarly non-classical;  at least close to the Page time (the half-life of the BH in units of entropy \cite{Page}) when their  complete sets of eigenvalues are  practically equal. It should still be true at earlier times provided that a  condensate (or some  other highly occupied state) is not ``hidden''  inside of  the BH.  We will, however, argue that such a state cannot be hidden after a time scale which is many orders of magnitude smaller than the Page time for a macroscopic BH.

We use the term geometry to mean the spacetime metric as well as its derivatives, such as the various curvature invariants. It should also be emphasized that, when talking about a non-classical interior and its lack of a semiclassical geometry (see Section~4), we are referring  to its description in terms of asymptotic fields (those of the Hawking modes) when expressed in the  standard Fock basis. An interior observer could avoid our conclusions but the corresponding choice of observables would not have a simple physical meaning from an asymptotic perspective. An asymptotic observer could likewise avoid our conclusions by averaging over suitably large distances, time scales and/or spans in frequency. The outcome  of such an averaging procedure would be the Schwarzschild metric.

The paper concludes with a brief overview in Section~5, followed by an appendix with some supplemental analysis.

\section{Evidence for a  non-classical state of Hawking radiation}

According to Page \cite{Page}, the state of the BH radiation starts to purify just after the Page time, which is the time when its Hilbert space and that of the BH interior are equal in size. Consequently, the radiation is in a highly entangled state at later times. According to Bell \cite{Bell}, the state of the Hawking radiation has then become non-classical, as one cannot assign a classical distribution function for such a highly entangled state.  We want to go further and show  that the radiation is in a non-classical state even well before the Page time. This is the primary goal of this section.

It will be shown by explicit calculation that, for a macroscopic BH, the occupation numbers of the Hawking radiation are small. This result is essentially contained in the early calculations of Hawking \cite{info} and, in particular, of Page \cite{Page}. However, since this is central to  our purpose, we will review and highlight the necessary ingredients. It will then be shown that these  sparsely occupied modes  imply large relative fluctuations in their occupation numbers. A consequence of having large relative fluctuations in a state is that a  mean-field approximation cannot be applied. In particular, large relative fluctuations in the stress--energy--momentum  (SEM) tensor imply similarly large relative fluctuations in the spacetime  curvature \cite{Jaekel}, which in turn means  that a semiclassical geometry cannot be trusted to  faithfully describe  the state of the Hawking radiation.

\subsection{Occupation numbers of Hawking radiation}

Let us start  by reviewing the discrete wave-packet basis for modes of Hawking radiation \cite{Hawk}. The discussion closely follows  that of \cite{density},  where additional details can be found.

Using  continuum normalization,  one can express the incoming and outgoing Hawking modes  as
                \begin{eqnarray}
f_{\omega lm}(v, r, \theta, \phi) &=& F_{\omega lm}(r) Y_l^m (\theta, \phi) e^{i \omega v} \;, \\
p_{\omega lm}(u, r, \theta, \phi) &=& P_{\omega lm}(r) Y_l^m (\theta, \phi) e^{i \omega u} \;,
\end{eqnarray}
where $l$ and $m$ are the angular-momentum eigenvalues, and $v$ and $u$ are, respectively, the advanced and retarded Eddington--Finkelstein coordinates.

A basis of wave packets can then be  defined as
                \begin{eqnarray}\label{PositiveFreq}
f_{jnlm} (r, \theta, \phi) &=& \epsilon^{-1/2} \int_{j \,\epsilon}^{(j+1) \,\epsilon} e^{-2\pi i \, n \, \omega/\epsilon} \,f_{\omega lm}(v, r, \theta, \phi) \, d\omega \;, \\
p_{jnlm} (r, \theta, \phi) &=& \epsilon^{-1/2} \int_{j \,\epsilon}^{(j+1) \,\epsilon} e^{-2\pi i \, n \, \omega/\epsilon} \,p_{\omega lm}(u, r, \theta, \phi) \, d\omega \;,\label{packet}
\end{eqnarray}
where $\epsilon$ is a dimensional ``resolution'' parameter, and the  continuous frequency $\omega$ and the null coordinates $u$, $v$ have been traded away for a pair of integers  $j \ge 0$ and $n$. The frequencies from which a  wave packet $f_{jnlm}$ or $p_{jnlm}$ is  built is localized in the range $\;j\, \epsilon \le \omega \le (j+1) \, \epsilon\;$. An incoming wave packet $f_{jnlm}$ is centered about $\;v = 2\pi n /\epsilon\;$, whereas an outgoing  $p_{jnlm}$  is centered about $\;u = 2\pi n /\epsilon\;$. The width of either type   is equal to $2\pi/\epsilon$.

The wave packets which are emitted from the BH during a certain period of time,
$\;t_0 \le t \le t_0 + \Delta t\;$, and then detected at some fixed distance away  will be  localized in the corresponding range of retarded time, $\;u_0 \le u \le u_0 + \Delta t\;$. For a long-enough time interval, the width $2\pi/\epsilon$ can be chosen such that many wave packets arrive during this same range of retarded time, $\;\epsilon\;\Delta t\gg 1$ and yet still have a good frequency resolution, $\;\epsilon \ll T_H\;$  ($T_H$ is the Hawking temperature). In this case, each wave packet can be treated as a monochromatic mode of some fixed frequency. And  so a sum over wave-packet position ({\em i.e.,} a sum over $n$)
within the interval $\;\Delta n = \Delta t\; \epsilon/(2 \pi)\;$ can be approximated by
$\;
\sum\limits_{n} \approx \frac{1}{2\pi}\Delta t\;\epsilon\;,
$
whereas a sum over discrete frequencies $\;j=\omega/\epsilon\;$ can be approximated by
$\;
\sum\limits_{j} \approx \int d\omega/\epsilon \;.
$
It follows that, in this approximation, the total number of modes which can be detected during a time interval $\Delta t$  does not depend on the choice of $\epsilon$,
\begin{equation}\label{integration}
\sum\limits_i\; =\; \sum\limits_{j}\sum\limits_{n}\;=\;\frac{1}{2\pi} \Delta t \, \int d \omega\;.
\end{equation}
This can be compared to the standard sum over modes for a thermal state in some restricted volume $V$ in $d$ space dimensions,
$\;\sum\limits_i=V\int d^d p/(2\pi)^d\;$.

The occupation numbers of the modes in the wave-packet basis are given by Hawking's famous calculation \cite{Hawk},
\begin{equation}\label{important}
\langle n_{jnlm} \rangle\;=\; \frac{\Gamma_{jnlm}}{e^{\frac{j\epsilon}{T_H}}-1}\;,
\end{equation}
where $\Gamma_{jnlm}$ are the   grey-body factors for BH emission.
The emission of  modes with large values of $l$ is highly suppressed, as is the emission of modes with masses in excess of the Hawking temperature. For macroscopic BHs, the Hawking temperature scales as $\;T_H\simeq 5\times 10^{-12}~{\rm eV}~\frac{M_\odot}{M_{BH}}\;$. The only modes that are light enough to be emitted from macroscopic BHs are then photons, gravitons and, possibly but unlikely, one species of neutrinos.~\footnote{A single very light neutrino species is still allowed by the currently available data, which mostly constrains the mass-square differences between the three known species.} The discussion will therefore be limited  to the emission of just photons and gravitons while the case of a massless neutrino is relegated to Appendix A.

According to Page's analysis of Schwarzschild BHs \footnote{ We restrict the following discussion to four-dimensional spacetime. The results can be adapted to higher dimensions and are found to be similar in nature. This claim is not in contradiction with \cite{hod} because here, unlike there, the focus
is on the occupation numbers of individual modes.} \cite{Page2}, the low-frequency, $\;\omega R_S \ll 1\;$, grey-body factors for massless spin-1 (photon) and spin-2 (graviton) modes are
($A_{H}$ is the BH horizon area)
\bea
\label{lessImp}
\Gamma(\omega=j\epsilon,s=1)&=&\frac{4A_{H}}{9\pi}M_{BH}^2 \omega^4\;, \\
\Gamma(\omega=j\epsilon,s=2)&=&\frac{16A_{H}}{225\pi}M_{BH}^4 \omega^6\;.
\eea
Hence, the grey-body factors are very small, which leads to very small occupation numbers. For example, substituting
$\;\omega R_S = 0.1\;$  into the above expressions and choosing $\;\ell=s\;$, one obtains the following occupation numbers,
\begin{align}
\langle n \rangle (l,\ s=1)\;\approx\; 1\times 10^{-5}\;,
\\
\langle n \rangle (l,\ s=2)\;\approx\; 7\times 10^{-9}\;.
\end{align}

In the high-frequency regime, $\;\omega R_S \gg 1\;$, the Hawking modes have enough energy to pass over the potential barrier; meaning that the grey-body factors become unity.  Nevertheless, the occupation numbers are exponentially suppressed by
the  Boltzmann factor,
$
\langle n \rangle (\omega R_S \gg 1 ) \;\approx\; e^{-4\pi R_S \omega} \;\ll\; 1
$.

To address the  intermediary cases, we will rely on  the numerical calculations of Gray and Visser  for the grey-body factors \cite{Visser}  (also, \cite{sparse}). In this intermediate-frequency regime, one finds that the occupation numbers of the photons are, at most, of order $10^{-2}$ and  those of the gravitons are smaller than about $10^{-3}$.

Our conclusion is that all of the non-vanishing  occupation numbers for single modes are very small. Of course, according to Eq.~(\ref{integration}), if the occupation numbers are integrated over a period of time that is much longer than $\;1/\omega\sim 1/T_H\;$, they will increase linearly with time and  eventually become large.  But, in spite of appearances, a typical mode will still have  occupancy below unity. This is because the number of occupied modes is similarly growing linearly in time \cite{density}. Also, if the BH can emit more massless species or, equivalently, happens to exist in a higher-dimensional space time, the occupation numbers can appear to become large. But, even so, the phase-space density of the occupied modes would remain small.

\subsection{Comparison to the radiation emitted by the Sun}

Suppose that the Sun is replaced by a hypothetical  BH with
 precisely  the same temperature --- if you didn't look
outside, could you feel the difference? The answer to this question is, perhaps surprisingly, yes!

Our  objective here is to  compare  the occupation numbers
of BH radiation  with those of the Sun (which is meant
to represent a typical black-body emitter).
The rate of emission of photons from a black body with a surface area of $A$ and a temperature of $T$ is
\be
\Gamma\;=\; \frac{\;\zeta(3)}{2\pi^2}T^3 A\;.
\ee
This equation is not strictly valid for BH emission because it neglects the grey-body factors. However, the inclusion of
these factors would only strengthen our conclusion.

It follows that the ratio of the emission rate  of the Sun to that of a  BH radiating at the same (solar) temperature
$T_{\odot}$ is
\begin{equation}
\frac{\Gamma_{\odot}}{\Gamma_{BH}} \;=\; \frac{A_{\odot}}{A_{BH}}\;.
\end{equation}
One can phrase this result as follows: For the same temperature, the area of the BH is much smaller than the area of the Sun, hence the emitted power is also much smaller.
Recalling that $\;R_S = (4\pi T_H)^{-1}\;$, the ratio is given by
\begin{equation}
\left(\frac{4\pi k_B T R_{\odot}}{\hbar c}\right)^2 \;\approx\; 5 \times 10^{32}\;.
\end{equation}
This ratio is discussed in \cite{Visser}.

This result is at the crux of why the occupation numbers of individual modes (and their fluctuations) can be resolved for a BH but not for a typical semi-classical emitter like the Sun. The relevant distinction between the BH and the Sun is in their  respective time-resolution scales, $\frac{1}{\Delta t} \sim \Gamma_{BH}\sim T_H\;$. However, for the Sun,  $ \frac{1}{\Delta t} \sim \Gamma_{\odot} \sim 10^{32} \; T_{\odot}\;$. The bottom line is that the individual modes in solar radiation can {\em never} be resolved as a strict matter of principle.

\subsection{Fluctuations}

We now want to determine the strength  of  the quantum fluctuations in  the occupation numbers for the state of BH radiation.

A generating function from \cite{density}
is quite  helpful with occupation-number calculations,
\begin{equation}
f(\lambda_i,\mu_j) \;=\; \langle 0_-| e^{\mu_i \hat{b} ^{\dagger}_i} e^{\lambda_i \hat{b}_i} |0_-\rangle\;.
\end{equation}
Here, just like in Hawking's work, the state $|0_-\rangle$ is the initial vacuum state of the gravitationally collapsing body,  $b_k$ is an annihilation operator for the radiation modes as seen by a far-away observer and $b_k ^{\dagger}$ is the corresponding creation operator. (The subscript $k$ is short for $j$, $n$, $l$, $m$.)

The prescription for calculating occupation numbers  is~\footnote{The order of differentiation is fixed by the normal ordering of operators. That is, that annihilation operators are on the right.}
\begin{equation}
\langle 0_-| b_k ^{\dagger} b_k | 0_-\rangle \;=\; \frac{\partial ^2 f(\lambda_i,\mu_j)}{\partial \lambda_k \partial \mu_k}|_{\lambda_i=\mu_j=0}\;=\;\langle n_k \rangle \;.
\end{equation}
And, by similar reasoning,
\begin{equation}
\langle 0_- |\left(b_k ^{\dagger} \right)^{2} b_k ^2|0_-\rangle \;=\; \frac{\partial^2}{\partial \mu_k^2 }\frac{\partial^2}{\partial \lambda_k^2 }f(\lambda,\mu) |_{\mu=\lambda=0} \;=\; 2\langle n_k \rangle ^2\;.
\end{equation}
On the other hand,
\begin{equation}
\langle n_k ^2\rangle\;=\;\langle b_k ^{\dagger} b_k b_k ^{\dagger} b_k\rangle \;=\; \langle b_k ^{\dagger} ( b_k ^{\dagger} b_k + 1) b_k\rangle\;=\;\langle n_k\rangle + 2\langle n_k\rangle^2\;.
\end{equation}
So that
\begin{equation}\label{OccNumFluc}
\Delta n_k^2 \;=\;
\langle n_k^2 \rangle - \langle n_k \rangle^2 \;=\;
\langle n_k\rangle + \langle n_k\rangle^2\;.
\end{equation}

Since  the average occupation numbers are small, $\;\Delta n_k^2\simeq \langle n_k\rangle\;$, it follows that the relative fluctuations are large,
\be \label{hier1}
\frac{\Delta n_k^2}{\langle n_k\rangle^2}\;\simeq \; \frac{1}{\langle n_k\rangle} \;\gg \;1\;.
\ee

\subsection{States with small occupation numbers are non-classical}
We are now finally positioned to demonstrate the non-classical nature of the BH radiation, following Chapter~8 of \cite{duncan}. To this end, let us consider one specific mode of radiation. Applying the polar-decomposition theorem to the annihilation and creation operators of the mode, one can formally define their phase,
\bea
\widehat{a}_k &=& e^{i\widehat{\Phi_k}}~ \sqrt{\widehat{n}_k} \;,\\
\widehat{a}_k^{\dagger}&=&\sqrt{\widehat{n}_k}~e^{-i\widehat{\Phi_k}}\;.
\eea
The standard commutation relation
$\;[\widehat{a}_k,\widehat{a}_k ^{\dagger}]=1\;$ can now be expressed as $\;[e^{i\widehat{\Phi_k}},\widehat{n}_k]=e^{i\widehat{\Phi_k}}\;$.
These formal expressions are, however, not quite precise. For one thing, the polar-decomposition theorem is not strictly valid for an infinite-dimensional Hilbert space. For another, the phase is defined modulo $2\pi$.

To overcome these difficulties, one can truncate the Hilbert space such that its new dimension $N$ is large but finite \cite{duncan}, and use the sine and cosine of the phase instead of the phase itself. For the truncated Hilbert space, the commutator becomes
$\; [e^{i\widehat{\Phi_k}},\widehat{n}_k]=e^{i\widehat{\Phi_k}}-n_{N}|n_{N}\rangle \langle n_N| \;$. The last term, which   is a consequence of the truncation, is insignificant because the occupation number of the cutoff
state $|n_{N}\rangle$ must be negligible for the truncation procedure to make sense. Then, approximately,
\begin{equation}
	[e^{i\widehat{\Phi_k}},\widehat{n}_k]\;\approx\; e^{i\widehat{\Phi_k}}\;,
\end{equation}
\begin{equation}
	[\widehat{n}_k, e^{-i\widehat{\Phi_k}}]\;\approx\; e^{-i\widehat{\Phi_k}}\;.
\end{equation}

The addition and subtraction of these (approximate) commutation
relations leads to
\begin{equation}
	[\widehat{n}_k,\sin(\widehat{\Phi_k})]\;=\;i\cos(\widehat{\Phi_k})\;,
\end{equation}
\begin{equation}
	[\widehat{n}_k,\cos(\widehat{\Phi_k})]\;=\;-i\sin(\widehat{\Phi_k})\;.
\end{equation}
These, in turn, imply uncertainty inequalities,
\begin{equation}
	\Delta n_k \Delta \cos(\Phi_k) \;\ge\; \frac{1}{2}|\langle \sin(\Phi_k) \rangle |\;,
	\label{uncertain1}
\end{equation}
\begin{equation}
	\Delta n_k \Delta \sin(\Phi_k)\;\ge\; \frac{1}{2}|\langle \cos(\Phi_k) \rangle|\;.
	\label{uncertain2}
\end{equation}
Using $\;|\langle \sin(\Phi_k)\rangle|^2 = \langle \sin(\Phi_k)^2 \rangle-\Delta \sin(\Phi_k)^2 \;$, we can sum the squares of Eqs.~(\ref{uncertain1}) and~(\ref{uncertain2}) to deduce that
\be
\left(\frac{1}{4}+\Delta n_k^2\right)  \left(\Delta \sin(\Phi_k)^2+\Delta \cos(\Phi_k)^2\right) \;\ge\; \frac{1}{4}\;.
\label{uncertain3}
\ee

Meanwhile, the classical limit is achieved when all of the following inequalities are satisfied,
\begin{eqnarray}
	\Delta n_k \;\ll\; \langle n_k \rangle\;,
	\label{class1}\\
	\Delta \sin(\Phi_k)\; \ll\; |\langle \sin(\Phi_k) \rangle |\;,
	\label{class2}\\
	\Delta \cos(\Phi_k) \;\ll \;|\langle \cos\Phi_k\rangle |\;.
	\label{class3}
\end{eqnarray}
Since $\;|\langle \sin(\Phi_k)\rangle|\le 1\;$, Eq.~(\ref{class2}) implies $\;\Delta \sin(\Phi_k) \ll 1\;$ and, similarly, Eq.~(\ref{class3}) implies that $\;\Delta \cos(\Phi_k) \ll 1\;$. This means that $\Delta \cos(\Phi_k)^2+\Delta \sin(\Phi_k)^2 \ll 1$. Combining this inequality with the one in Eq.~(\ref{uncertain3}), we arrive at $\;\Delta n_k \gg 1\;$. But then, from Eq.~(\ref{class1}), it must follow that
\begin{equation}
	\langle n_k \rangle \;\gg\; 1\;.
\end{equation}

Consequently, for the state of a  single mode of the BH radiation to have a description as a classical field, it must have a large occupation number. But we already know from Subsection~2.1 that the occupation numbers are small. Hence, the state of the radiation must be non-classical; the photons cannot be described in terms of classical (electromagnetic) fields and similarly for the gravitons. Averaging over time or space or a number of different modes will, of course, alter the result.

\subsection{Large relative fluctuations in the stress-energy-momentum tensor}
We will again focus on one particular mode of radiation $\;k=(j,n,l,m)\;$. The associated field operator for an outgoing wave takes the form
\begin{equation}\label{Decomposition}
	\Phi_k (u,r,\theta,\phi) \;=\; p_k (u,r,\theta,\phi) b^{\dagger} _k \;+\; p^{\ast}_k (u,r,\theta,\phi) b_k\;,
\end{equation}
where the positive-energy component $p_k(u,r,\theta,\phi)$
was defined in Eq.~(\ref{packet}). Let us reemphasize that
this component, the creation operator $b^{\dagger} _k$
and their respective conjugate and adjoint are those as seen by an observer far away from the BH.

We next want to construct the normal-ordered SEM tensor $:T_{\mu\nu}:$
for the field
$\Phi_k$. Our eventual goal
is to compute the strength of the relative fluctuations for this tensor,
$\frac{\Delta :T_{\mu \nu}: ^2}{\langle :T_{\mu \nu}: \rangle^2}$.
We will discuss the SEM tensor for a massless scalar field, as using vectors and tensors would not affect our conclusions but would clutter up the presentation,
\begin{equation}
T_{\mu \nu}\;=\;\partial_{\mu} \Phi_k \partial_{\nu} \Phi_k-\frac{1}{2} g_{\mu \nu} \partial_{\lambda}\Phi_k \partial ^{\lambda} \Phi_k\;.
\label{tmunu1}
\end{equation}
The substitution of Eq.~(\ref{Decomposition}) into Eq.~(\ref{tmunu1}) yields
\be
T_{\mu \nu } \;=\; t_{\mu \nu} \left(b_k ^{\dagger}\right) ^2 + t_{\mu \nu} ^* b_k ^2 +\beta_{\mu \nu} b_k ^{\dagger} b_k + \beta_{\mu \nu} ^* b_k b_k ^{\dagger}\;.
\ee
Normal ordering and using the definition of the number operator,
\begin{equation}
	:T_{\mu \nu}:\;=\;t_{\mu \nu} \left(b_k ^{\dagger}\right) ^2 + t_{\mu \nu} ^* b_k ^2 +2\text{Re}(\beta_{\mu \nu}) n_k\;,
\end{equation}
where we have defined
\begin{eqnarray}
	t_{\mu \nu}& =& \partial_{\mu} p_k \partial_{\nu} p_k - \frac{1}{2}g_{\mu \nu} \partial_{\lambda}p_k \partial^{\lambda}p_k \;,\\
	\beta_{\mu \nu}&=&\partial_{\mu} p_k \partial_{\nu} p_k ^* - \frac{1}{2}g_{\mu \nu} \partial_{\lambda}p_k \partial^{\lambda}p_k ^*\;.
\end{eqnarray}

The mean value of the normal-ordered SEM tensor is
then  given by
\begin{equation}
	\langle 0_-|:\!T_{\mu \nu}\!:|0_-\rangle \;=\; 2\text{Re}(\beta_{\mu\nu})\langle n_k\rangle\;.
\end{equation}
Here, we have used that
the mean values of $b_k^2$ and $\left(b_k ^{\dagger}\right)^2 $ vanish. This was shown explicitly by Hawking \cite{Hawk}, who used a Bogoliubov transformation to show that either pair of operators
annihilates the initial vacuum state.
To determine the fluctuations, one needs to first calculate the square of the normal-ordered SEM tensor. It contains terms such as $b_k ^4$, $b_k ^{\dagger} b_k ^3$ with vanishing expectation values, which will be denoted by an ellipsis, and the following terms with non-zero expectation values:
\begin{equation}
	(:\!T_{\mu \nu}\!:)^2 \;=\;4\text{Re}(\beta_{\mu\nu})^2 \langle n_k\rangle^2
+|t_{\mu\nu}|^2 \left( \left(b_k ^{\dagger}\right)^2 b_k ^2+b_k ^2 \left(b_k ^{\dagger}\right)^2\right) \;+ \;[\cdots]\;.
\label{teesquared}
\end{equation}

To proceed further, we will call upon some identities that follow
from the generating function of Subsection~2.3,
\bea
\langle n_k ^2 \rangle &=& \Delta n_k ^2 + \langle n_k \rangle ^2\;, \\
\langle 0_-|\left(b_k ^{\dagger} \right)^2 b_k ^2 |0_-\rangle &=& \Delta n_k ^2 + \langle n_k \rangle^2 - \langle n_k\rangle\;, \\
\langle 0_-| b_k ^2 \left( b_k ^{\dagger}\right)^2 |0_-\rangle &=& 2+3\langle n_k\rangle +\langle n_k \rangle ^2 + \Delta n_k ^2\;,
\eea
which  then leads to
\bea
\langle 0_-| (:\!T_{\mu\nu}\!:)^2 |0_-\rangle \;=\; \left( 4\text{Re}(\beta_{\mu \nu})^2 +2|t_{\mu\nu}|^2 \right)\Delta n_k ^2 &+& \left( 4\text{Re}(\beta_{\mu \nu})^2 +2|t_{\mu\nu}|^2 \right)\langle n_k \rangle^2 \nonumber \\ &+& 2|t_{\mu\nu}|^2 \langle n_k \rangle +2|t_{\mu\nu}|^2\;.
\eea

The above results allow one to calculate the variance and then
the relative fluctuation strength of the normal-ordered SEM tensor. The former is
\begin{equation}
	\Delta :\!T_{\mu\nu}\!: ^2 \;=\;(4\text{Re}(\beta_{\mu\nu})^2 \;+\; 2|t_{\mu\nu}|^2)\Delta n_k ^2\; + \;2|t_{\mu\nu}|^2 \left(\langle n_k \rangle^2+\langle n_k\rangle +1 \right) \;.
\end{equation}
Using equation (\ref{OccNumFluc}) and dividing
the above by the square of the mean value, we have for the latter
\begin{equation}\label{RelativeStrEnergyFluc}
	\frac{\Delta :\!T_{\mu\nu}\!:^2}{\langle :\!T_{\mu \nu}\!:\rangle^2}\;=\;\left(1+ \frac{|t_{\mu\nu}|^2}{ \text{Re}(\beta_{\mu\nu})^2}\right) \frac{\Delta n_k ^2}{\langle n_k\rangle^2}\;+\;\frac{|t_{\mu\nu}|^2}{2\text{Re}(\beta_{\mu\nu})^2} \frac{1}{\langle n_k \rangle^2}\;,
\end{equation}
which implies the  bound
\begin{equation}
\label{SEMfluc}
\frac{\Delta :\!T_{\mu\nu}\!: ^2}{\langle :\!T_{\mu \nu}\!:\rangle^2}\;\geq\; \frac{\Delta n_k ^2}{\langle n_k \rangle ^2}\;.
\end{equation}

Recalling Eq.~(\ref{hier1}), one can see that the right-hand side of the bound~(\ref{SEMfluc}) is much greater than unity. It can now be concluded that, for individual modes of
Hawking radiation, the relative fluctuations of their associated SEM tensors are  large
\begin{equation}
\label{SEMfluc1}
\frac{\Delta :\!T_{\mu\nu}\!: ^2}{\langle :\!T_{\mu \nu}\!:\rangle^2}\gg 1\;.
\end{equation}

\subsection{The state of black hole radiation is not semi-classical}
Another way to determine whether a state is non-classical is to study its Wigner function. This is
because the Wigner function allows one to calculate quantum expectation values in a way that closely resembles
the calculation of classical averages ({\em e.g.}, \cite{Wigner2}),
\begin{equation}\label{SuperWigner}
	\langle A \rangle \;=\; \int\int \frac{dqdp}{2\pi} W(q,p)A(q,p)\;,
\end{equation}
where $q$, $p$ are a pair of canonical conjugate variables, $A(q,p)$ is the Wigner--Weyl representation of some operator and $W(q,p)$ is the Wigner function.
For a state whose density matrix is
${\widehat \rho}$, the latter function is expressible as
\begin{equation}
	W(q,p)\;=\;\frac{1}{\pi \hbar} \int_{-\infty}^{\infty} \langle q+y |\widehat{\rho}|q-y\rangle e^{2ipy} dy\;.
\end{equation}

The common diagnostic, in this context, for identifying a non-classical state
is to look for regions of phase space where the Wigner function becomes negative. This is because the Wigner function in Eq.~(\ref{SuperWigner}) plays the same role as a classical probability distribution function would. We will, however, be following a different route and show that a semiclassical expansion of the Wigner function breaks down for a thermal state with small-enough occupation numbers. This means that one cannot find a classical probability distribution function that can approximate the quantum distribution function.~\footnote{Due to the
omission of the grey-body factors in this subsection,
the failure is even more dramatic than what will be shown
here.}

Let us now specialize to the outgoing component for some particular photon mode. The conjugates $p$ and $q$ can then be related to its associated electromagnetic field. In terms of the wave-packet
basis of Subsection~2.1, these relations can be expressed as
\begin{equation}
	\overrightarrow{\widehat{E}}_{jnlm} (u,r,\theta,\phi)\;=\;\omega \overrightarrow{\widehat{q}}(u) R_{jnl} (r) Y_{lm} (\theta,\phi)\;,
\end{equation}
\begin{equation}
	\overrightarrow{\widehat{B}}_{jnlm} (u,r,\theta,\phi)\;=\;\overrightarrow{\widehat{p}}(u) \tilde{R} _{jnl} (r) Y_{lm} (\theta,\phi)\;.
\end{equation}
The labels are the same as before, but it should be emphasized that
$\;j=\omega/\epsilon\;$, the radial functions are orthogonal with respect to $j$ and the quantum number $n$ is not an occupation number.
Let us further specify that this is for a single mode of {\em thermal} radiation and that all degrees of freedom besides $j$ have been left implied. Then the single-mode density matrix will have the
standard form,~\footnote{The total density matrix is
	$\;\widehat{\rho}_{tot}=\Pi_j \widehat{\rho}_j\;$. This separation is  possible as long as there are no interactions between
	the modes.}
\begin{equation}
	\widehat{\rho}_{j} \;=\; \frac{1}{Z_j} e^{-\Omega_j \widehat{n_j}} \quad , \quad \Omega_j\;=\;\frac{\hbar \omega_j}{T} \quad, \quad Z_j\;=\; {\rm Tr}[\widehat{\rho}_j] \;.
\end{equation}

The Wigner function for a single mode of thermal radiation is given by \cite{Wigner2}
\begin{equation}
	W(q,p)\;=\;\frac{1}{\hbar\pi}\tanh\left(\frac{\hbar \omega_j}{2T} \right)
	\exp\left( -\frac{1}{\hbar\omega_j}\tanh\left(\frac{\hbar \omega_j}{2T} \right) \left(p^2+\omega_j^2 q^2\right) \right)\;.
\end{equation}
Expanding this expression in terms of  $\frac{\hbar \omega_j}{T}$ ({\em i.e.}, the dimensionless $\hbar$), we then have
\begin{equation}
	W(q,p)\;=\;\left(\frac{\omega_j}{2\pi T}-\frac{\hbar^2 \omega_j^3}{24\pi T^3}+O(\hbar^4)\right)\exp\left(-\left(\frac{1}{T}-\frac{\hbar^2 \omega_j^2}{8T^3}+O(\hbar^4) \right) \frac{1}{2}(p^2+\omega_j^2 q^2) \right)\;. \label{Wexpand}
\end{equation}

In the zeroth-order approximation, the expansion reduces to
\begin{equation}
	W(q,p) \;=\; \frac{\omega_j}{2\pi T} e^{-\frac{1}{2T}(p^2+\omega_j^2 q^2)}\;,
\end{equation}
which  can be viewed as a classical distribution function for a system of photons at temperature $T$. Hence, the terms which are of higher order in $\frac{\hbar \omega_j}{T}$ can be interpreted as small quantum corrections to the leading terms. We can conclude that the  semiclassical expansion of the Wigner function is in terms of the dimensionless $\hbar$, $\frac{\hbar \omega_j}{T}$.

The Wigner expansion~(\ref{Wexpand}) can also be expressed in terms of the mode's occupation number. Using
the expression for a thermal occupation number,
\begin{equation}
	\label{thermalOccupation}
	\langle n_j\rangle \;=\; \frac{1}{e^{\frac{\hbar \omega_j}{T}}-1} \;,
\end{equation}
one finds that
\be
\label{thermalOccupation1}
\frac{\hbar \omega_j}{T}\;=\;\ln\left(1+\frac{1}{\langle n_j\rangle}\right)\;.
\ee

It can now be observed that a small dimensionless $\hbar$ corresponds to a small value of $\;1/\langle n_j\rangle\;$. So that, in terms of an expansion in occupation number, the dimensionless $\hbar$ is equal to $1/\langle n_j\rangle$. A semiclassical expansion thus requires $\;\langle n_j\rangle \gg 1\;$, which is  certainly not satisfied  for a mode of Hawking radiation. We can therefore conclude,  once again, that the quantum corrections
are too large for the Hawking radiation to be viewed as a semiclassical state.

\section{A non-semiclassical geometry}
Next, we consider an observer who is far away from the BH ($r\gg R_S$) and collecting the Hawking particles for one particular mode or, more realistically, for a narrow band of frequencies covering a small fraction of the modes. It will be shown in what follows that the relative quantum fluctuations in the curvature induced by the specific modes are large for such an observer,
even though the average curvature is very small.

\subsection{Large relative curvature fluctuations }

The expectation value of the curvature can be obtained from the semiclassical version of the Einstein equation,
\begin{equation}\label{Einstein}
	\langle (G_{\mu \nu})_k \rangle \;=\;  \frac{8\pi\hbar }{M_p ^2} \langle (T_{\mu \nu})_k \rangle\;,
\end{equation}
but then what about the fluctuations of these tensors?

To address this question, let us first consider the spacetime metric. It is safe to assume that linearized gravity is valid in this observer's (approximately flat) region of spacetime; hence,
\begin{equation}
	g_{\mu \nu}\;=\; \eta_{\mu \nu} + h_{\mu \nu}\;,
\end{equation}
where $h_{\mu\nu}$ is a perturbation of the Minkowski metric $\eta_{\mu\nu}$.
By construction, the quantum field $h_{\mu \nu}$ includes not only the effects of the Hawking particles but also the (small) deviation between the Schwarzschild metric and flat spacetime. We can then use the results of \cite{Jaekel}, whose derivation is reviewed in Appendix B,   that the fluctuations in the curvature are equal to the fluctuations in the SEM tensor.
  \begin{equation}
  \label{JaekelEquation}
	\Delta (G_{\mu \nu})_k ^2 \;=\; \left(\frac{8\pi \hbar}{M_p ^2}\right)^2 \Delta (T_{\mu \nu})_k ^2 \;,
\end{equation}
which combines with Eq.~(\ref{Einstein}) to give
\begin{equation}
	\frac{\Delta (G_{\mu \nu})_k ^2}{\langle (G_{\mu \nu})_k \rangle^2 } \;=\; \frac{\Delta (T_{\mu \nu})_k ^2}{\langle (T_{\mu \nu})_k \rangle ^2 }\;.
\end{equation}

Using the last equation and the inequalities in~(\ref{hier1}) and~(\ref{SEMfluc}), we now know that
\begin{equation}
	\frac{\Delta (G_{\mu \nu})_k ^2}{\langle (G_{\mu \nu})_k \rangle ^2} \; \geq \frac{\Delta n_k ^2}{\langle n_k \rangle ^2} \;\gg\; 1\;.
	\label{curvfluc}
\end{equation}
It can therefore be concluded that a single mode of Hawking radiation induces large relative quantum fluctuations in the Einstein tensor for the same value of $k$.  Of course, this conclusion can easily  be missed if one averages over large time scales or regions of space.

An interesting manifestation of the fact that there are no ``one-size-fits-all'' criteria for distinguishing between non-classical and semiclassical states is the following.
The relation between the Ricci tensor and the gravitational field in the harmonic gauge is
 \begin{equation}
 \Box h_{\mu \nu} (x)\sim R_{\mu \nu}(x),
 \end{equation}
and so
 \begin{equation}
 h_{\mu \nu} \sim \int \frac{d^4 k}{(2\pi )^4} \dfrac{1}{k^2} R_{\mu \nu}(k) e^{i k\cdot x}.
 \end{equation}
Our argument is that, for specific values of $k$, $R_{\mu \nu}(k)$ has large fluctuations.
By averaging over a  range of $k$ values, one does get a quantity that has small fluctuations and therefore can be treated semiclassically. The distance between two spacetime points, which is proportional to yet another integral over $h_{\mu\nu}$,  has even smaller fluctuations.

On the one hand, Hawking quanta do not cause large relative quantum fluctuations for quantities giving some  notion of ``distance'' in classical gravity but, on the other hand, these quanta do induce large relative  fluctuations for some curvature momentum components. So, an observer could also decide to ignore the effects of individual Hawking modes by  averaging over sufficiently large spatial and/or temporal scales, as well as over a large  band of frequencies. The strength of  the {\em relative} fluctuations will then decrease accordingly. Such an averaging procedure amounts to tracing over the non-classical ``hair'', and what remains is a description of the geometry in terms of the classical Schwarzschild metric.

\subsection{The dimensionless $\hbar$ for the Hawking radiation versus the standard dimensionless $\hbar$ for gravity}

The inequalities in~(\ref{SEMfluc}) and ~(\ref{curvfluc}) suggest that
the dimensionless $\hbar$ for the Hawking radiation is
$\; \frac{\Delta n_k^2}{\langle n_k \rangle^2} \approx \frac{1}{\langle n_k \rangle }\;$. But the standard dimensionless $\hbar$ for a (massless) mode with momentum $k$ in semiclassical gravity is $\frac{k^2}{M_P ^2}$ \cite{Donaghue,Burgess}. According to this identification,
 quantum-gravity effects become important only at Planckian  energies. However, it has been argued here that, in the case of Hawking radiation, quantum effects are important at much lower energies. We would like to explain the reason for this difference in scales. As a first step in this direction, let  us review  the rationale that led to the identification of  $\frac{k^2}{M_P^2}$ as the dimensionless $\hbar$ in quantum  gravity.
 This will be done in a  way that makes the comparison between the two dimensionless  $\hbar$'s easier.

One starts by expanding the metric in terms of a rescaled      graviton $h_{\mu\nu}$,
\begin{equation}
g_{\mu \nu}\;=\; \eta_{\mu \nu} +\frac{1}{M_P} h_{\mu \nu}\;.
\end{equation}
The Planck mass $\;M_P=\sqrt{\hbar/G_N}\;$ is used  here because it is the only energy scale of the problem and, moreover, it leads to the canonical form of the kinetic term.

 Next, one imposes  the harmonic gauge,
$\;
	2k^{\mu} h_{\mu \nu} = k_{\nu} \eta^{\alpha \beta} h_{\alpha \beta}
\;$,
to obtain the Fourier-space representation of the
 normal-ordered,  linearized Ricci curvature tensor,
\begin{equation}
	:\widehat{R}_{\mu \nu}: \;=\; \frac{k^2}{2M_P} \widehat{h}_{\mu \nu}\;.
\end{equation}
Since  $\langle \widehat{h}_{\mu \nu}\rangle$ vanishes, the mean value of the curvature also vanishes,
\begin{equation}
	\langle :\widehat{R}_{\mu\nu}: \rangle\; =\;0\;.
	\label{expR}
\end{equation}

The quantum fluctuations can be determined by evaluating the variance of the curvature,
\begin{equation}
	\Delta :R_{\mu \nu}:^2 \;=\; \langle :\widehat{R}_{\mu \nu}:^2\rangle
	\;=\;\frac{k^2}{4M_P^2} k^2 \langle \widehat{h}_{\mu \nu}  \widehat{h}_{\mu \nu}: \rangle\;.
\end{equation}
This will not generally vanish in spite of its normal  ordering;
 {\em cf}, Eq.~(\ref{teesquared}).

Now, because of the vanishing expectation value in Eq.~(\ref{expR}), the relative fluctuation strength is an ill-defined
 quantity. It is, however, clear that the non-vanishing moments
of curvature will come in even powers of the Ricci tensor (or, more generally, the Riemann tensor), and so   the expansion must be in terms of $\frac{k^2}{M_P^2} k^2$. This identifies the dimensionless $\hbar$ with the ratio $\frac{k^2}{M_P^2}$ as expected.

Thus, one concludes that  quantum-gravity effects become important when $\;k^2 \approx M_p ^2\;$. However, when the dimensionless $\hbar$ is $\frac{\Delta n_k ^2}{\langle n_k\rangle ^2}$, the fluctuations are strong for small $\langle n_k\rangle$ and any value of $k$.  If relative curvature  fluctuations cannot be defined, one is forced to introduce the Planck mass, being the only available scale. But if they  can, the Planck mass cancels out because the relevant ratio  is between two quantities that each contains the same power  of $M_P$.

\section{The state of the BH interior}

We will now proceed to argue that the  non-classicality of the Hawking radiation ---  the large relative quantum fluctuations in the spacetime curvature for the individual modes, as shown in the last Section --- also  applies to the state of the BH interior.  If the interior curvature  is indeed fluctuating in this way, then Einstein's classical theory of gravity and the notion of a classical geometry  fail to be applicable in this region of spacetime.  See the end of Section~1 for the precise meaning of a non-classical interior and its lack of a geometric  description  in the current context.

Our argument relies on the BH interior being the purifier of  the emitted radiation (and {\em vice versa}).  How is it possible that such a large system can be approximately described as a pure state?
To be concrete, let us consider a solar-mass worth of collapsing matter. The collapsing matter has an  entropy on  the order of the number of baryon constituents, $\;S_{collapse}\approx \frac{M_{\odot}}{M_{proton}}\approx 10^{57}\;$. Since the collapse is a unitary process, the initial entropy is also approximately the final entropy. However, BHs of the same mass possess a Bekenstein--Hawking entropy on the order of $\;S_{BH}\approx 10^{77}$\;. Hence, the final state of the BH occupies  a fraction of order $10^{-20}$ in the phase space of possible states.    Such purity is probably the highest that one could contemplate in a real physical system.

Now, since the process of BH radiation is also unitary, the total density matrix $\rho_{tot}$ must remain pure throughout the BH's evaporation. Then, the two reduced density matrices, $\;\rho_{RAD}= Tr_{BH} [\rho_{tot}]\;$ and $\;\rho_{BH}=Tr_{RAD}[\rho_{tot}]\;$, are each other's purifying state. This  relationship implies that $\rho_{BH}$ and $\rho_{RAD}$  must share a common set of non-vanishing eigenvalues, as will be made explicit below. In particular, close to the Page time, the two reduced density matrices are equal because their dimensionality is (approximately) equal, and so they must have the same number of vanishing eigenvalues as well. It follows that the occupation numbers  in the BH interior must be equal to those of the external radiation in the same basis; meaning that the state of the BH interior is similarly non-classical. Our conclusion is that, due to strong relative quantum fluctuations, the BH interior cannot be described faithfully by a semiclassical geometry.

However, one possible way of evading our conclusion  might be if a highly-occupied state --- a condensate --- is hidden inside the  BH. This could no longer be an issue  after  the Page time, which is when the eigenvalues of the two subsystems can  be identified. Nevertheless,  we would like to make a stronger statement and will proceed accordingly.

Being a pure state, the total system  can be described by a single state $|\psi \rangle$.  It follows that, if the basis state  of the BH is denoted by $|a_i\rangle$ and that of the radiation is denoted $|b_j\rangle$, the total state can be expressed as
\begin{equation}
| \psi\rangle \;=\; \sum_{ij} A_{ij} |a_i\rangle_{BH} |b_j\rangle_{RAD}\;.
\end{equation}
Then, after a Schmidt decomposition,
\begin{equation}
|\psi\rangle \;=\; \sum_i A_i |c_i\rangle _{BH} |d_i \rangle _{RAD}\;,
\end{equation}
the reduced density matrices of the BH and the radiation are given by
\bea
\rho_{BH} & = & \sum_i |A_i|^2 |d_i\rangle \langle d_i|\;, \\
\rho_{RAD} & = & \sum_k |A_k | ^2 |c_k\rangle \langle c_k| \;,
\eea
where only the non-vanishing eigenvalues appear in the sums.
Since the state of the radiation is essentially thermal, we also know that
\begin{equation}
|A_k|^2 \;=\; \frac{1}{Z} e^{-\Omega_k n_k }\;.
\end{equation}

Let us now assume that, after the Schmidt decomposition, the sum includes a term of the form
$ A_0|0\rangle_{RAD} |n_C\rangle_{BH}$; that is, a ``hidden condensate." It then follows that
\begin{equation}
|A_0|^2 \;=\; \frac{1}{Z}\;.
\end{equation}
And, from the condition of normalization, it can be shown that \cite{density}
\begin{equation}
Z\;=\;\text{det}(1+\widehat{n})\;=\;e^{\ln\left( \text{det}(1+\widehat{n}) \right)}\;=\;e^{\text{tr}\left(\ln(1+\widehat{n})\right)}\;,
\end{equation}
which leads to
\begin{equation}
\ln(Z)\;=\;\frac{\Delta t}{2\pi} \int_0 ^{\infty} d\omega \ln\left(1+\frac{\Gamma (\omega)}{e^{\frac{\omega}{T}}-1} \right)\;=\;\frac{T\Delta t}{2\pi}\int_{0}^{\infty} dx \ln\left(1+\frac{\Gamma(x)}{e^x-1} \right)\;.
\end{equation}

The previous expression can be related to the average number of emitted particles,
\begin{equation}
N\;=\; \sum_i \langle n_i \rangle\; =\; \frac{T\Delta t}{2\pi} \int_{0} ^{\infty} dx \frac{\Gamma(x)}{e^{x}-1}\;,
\end{equation}
giving
\begin{equation}
\ln (Z) \;=\; N \frac{\int_{0} ^{\infty} dx\ln\left( 1+\frac{\Gamma(x)}{e^{x}-1}\right)}{\int_{0} ^{\infty} dx \frac{\Gamma(x)}{e^{x}-1}}\;.
\end{equation}

We can use the fact that $\;\Gamma(x)\ll 1\;$ for $\;x\ll 1 \;$   to approximate both integrals by integrating over $\;x>1$\; only. In this regime, photons and gravitons have a grey-body factor of unity and the logarithm can be expanded to first order.  Therefore,
\begin{equation}
\ln (Z) \;\approx\; N\;,
\end{equation}
meaning that
\begin{equation}
|A_0|^2 \;=\;\frac{1}{Z}\;\approx\; e^{-N}\;.
\end{equation}

The takeaway  of all this is that a condensate can indeed ``hide'' but only  until a small number of photons have  been emitted. For a solar-mass BH, this might as well be zero photons.

\section{Conclusion}

We have argued, under the assumption of  unitary evolution, that the state of the BH interior is non-classical because it is the purifier of a non-classical  state; namely, that of the Hawking radiation. The Hawking radiation is itself non-classical because it is a thermal-like state with small occupation numbers.   Further,  the non-classicality of the radiation implies that the induced geometry is also non-classical due to the large relative quantum fluctuations of the  spacetime curvature. This suggests that the state of the BH interior, which is similarly non-classical, is devoid of a meaningful description in terms of a semiclassical geometry.

It should be stressed that not all measurements would reveal the non-classical nature of the Hawking radiation or the BH interior. In fact, it might be that only a small fraction of   sufficiently precise  experiments would be useful in this   regard, as we would expect that BHs   maintain their ``baldness'' under most circumstances.    Nevertheless, this evasiveness is part and parcel when diagnosing states for signs of  non-classical behavior.

 Meanwhile, the notion that only a subset of observers or observations would be able to probe the BH interior has been gaining momentum in the literature ({\em e.g.}, \cite{Mathur,Raju}).
Although the details can differ from one study to the next,
one basic theme persists: Some type of averaging or coarse-graining procedure will inevitably lead to the standard picture of a  semiclassical  BH with an opaque horizon. This is  consistent with our results, which imply that the large fluctuations in the occupation numbers depended on looking
at single modes of radiation over relatively short time scales.

It is unclear what a non-geometrical interior really means. Given that entropy bounds do not permit the interior mass to collapse into a singular core, we have a contradiction with the Chandrasekhar limit. This suggests that all conventional matter, including a ``firewall'' of Hawking particles outside the horizon \cite{Sunny,Braun,Amps}, would inevitably collapse in this way. One must then look for exotic matter that would not be subject to collapse but, rather, would be sustained by quantum effects.  Elsewhere, we have suggested that the BH is filled with interacting, highly excited, long, closed strings \cite{strungout,emerge}.

\section*{Acknowledgments}
We would like to thank Doron Cohen and Sunny Itzhaki for valuable discussions. The research of RB and YZ was supported by the Israel Science Foundation grant no. 1294/16.
The research of AJMM received support from an NRF Incentive Funding Grant 85353, an NRF Competitive Programme Grant 93595 and Rhodes Research Discretionary Grants. AJMM thanks Ben Gurion University for their hospitality during his visit.

\appendix

\section{Small occupation numbers for neutrino emissions}

An interesting aspect of Page's numerical calculations in \cite{Page2} is that, if a BH has a mass of $\;M_{BH} > 10^{17}$~g~, then $81 \% $   of the emitted radiation is in the form of a massless or very light  neutrino (if one exists). Observations of neutrino oscillations, which allow for the  existence of one such neutrino, suggest that this possibility should not be completely excluded.  In this part of the Appendix, we consider the case of neutrino emissions and argue that their occupation numbers are also small, just like  for the emitted photons and gravitons.

A hypothetical massless neutrino has an occupation number of
the form
\begin{equation}
\langle n \rangle \left(j,n,l=s=\frac{1}{2}\right)\;=\;\frac{\Gamma(\omega R_S)}{e^{4\pi \omega R_S}+1}\;,
\label{occnu}
\end{equation}
and it has a  grey-body factor  at low frequencies, $\;\omega R_S \ll 1\;$, of \cite{Page}
\be
\Gamma \left(\omega=j\epsilon,s=1/2 \right)\;=\;M_{BH}^2 \omega^2\;.
\label{evenlessImp}
\ee
Then, for example,
$\;
\langle n \rangle \left(\omega R_S =0.1, l=s=\frac{1}{2}\right)\approx 5 \times 10^{-4}
\;$.
Meanwhile, when the frequencies are large, $\;\omega R_S \gg 1\;$, the grey-body factor approaches unity but then the occupation numbers are  exponentially (Boltzmann) suppressed,
$\;
\langle n \rangle \left(\omega R_S \gg 1, l=s=\frac{1}{2}\right)\approx e^{-4\pi \omega R_S } \ll 1
\;$.

A conservative estimate  for an upper bound on the neutrino occupation number goes  as follows:
\begin{equation}
\langle n \rangle \left(j,n,l=s=\frac{1}{2}\right)\;< \;\frac{1}{e^{4\pi \times 0.1}+1} \;\simeq\; 0.2\;.
\label{upperbound}
\end{equation}
Here, for $\;\omega R_S > 0.1\;$, we have used the maximum value of the grey-body factor and the minimal value for $e^{4\pi \omega R_S}$. Whereas, for $\;\omega R_S <0.1\;$, Page's expression for a low-frequency grey-body factor becomes valid. We do, however, expect the actual occupation numbers to be much smaller than the estimate in Eq.~(\ref{upperbound}).

\section{Relating the fluctuations of the Einstein tensor to the fluctuations  of the stress-energy-momentum tensor}

The semiclassical form of the Einstein equations relates the expectation value of the Einstein tensor to that of the SEM tensor,
\begin{equation}
\langle G_{\mu \nu} \rangle \;=\; \kappa \langle T_{\mu \nu} \rangle\;,
\end{equation}
where $\;\kappa=8\pi G\;$.
Our purpose here is to argue that  the fluctuations of the two tensors are similarly related,
\begin{equation}
\Delta G_{\mu \nu}^2 \;=\; \kappa^2 \Delta T_{\mu \nu}^2\;.
\label{tsem1}
\end{equation}

Jaekel and Reynaud (JR) \cite{Jaekel} have already shown that, for suitably weak gravity and low-enough energies, Eq.~(\ref{tsem1}) is
indeed valid in an approach that relied on the
 fluctuation-response relation from statistical mechanics. In the following, we review their analysis to make our own paper self-contained. This will include a review of the fluctuation-response relation and then a discussion on how to  apply it in the weak-gravity, low-energy
regime.

\subsection{Linear response formalism}
In the JR treatment, a correlation function in spacetime for two observables $A$, $B$ is defined by
\begin{equation}
C_{AB} (x)\; = \;\langle A(x) B(0) \rangle - \langle A(x) \rangle \langle B(0)\rangle\;,
\label{abovecorrelationfunction}
\end{equation}
where all expectation values are with regard to  free fields in the vacuum and in flat spacetime.
The symmetrized form of this correlation function is
\begin{equation}
\label{correlationFunction}
\sigma_{AB} (x)\; =\; \frac{1}{2\hbar} \left( C_{AB} (x) + C_{BA} (-x) \right)\;.
\end{equation}
What will eventually be needed here is $\;\sigma_{AA}(x) = \frac{1}{2\hbar} \left( C_{AA} (x) + C_{AA} (-x) \right)\;$. For a translation- and rotation-invariant theory, the correlation functions depend only on $|x|$; hence,
\begin{equation}
\label{correlationFunction1}
\sigma_{AA} (x)\; =\; \frac{1}{\hbar}  C_{AA} (x)\;.
\end{equation}

Let us next consider the susceptibility function $\chi_{AB} (k)$, which describes the response of an observable $A$ to a linear perturbation in another observable $B$,
\begin{equation}
A(k)\;=\;\chi_{AB} (k) B(k)\;.
\label{abovesigmachi1}
\end{equation}
The fluctuation-response relation then asserts that
\begin{equation}
\sigma_{AA}  (k)\;=\; \text{Im} (\chi_{AA} (k))\;.
\label{sigmachi1}
\end{equation}
The main idea of JR is that, from Eq.~(\ref{sigmachi1}), one can infer the fluctuations of the Einstein tensor by having knowledge about  $\chi$. So that, to this end, only one-point functions are  required.
	
\subsection{Proper fluctuations}

Proper fluctuations are  those which are self-induced. Following JR, we will label these with a superscript of $in$, meaning ``input''.

Let us now discuss linearized gravity in $D$ dimensions of spacetime, for which the metric adopts the familiar form
\begin{equation}
g_{\mu \nu}\;=\;\eta_{\mu \nu}+h_{\mu \nu}\;.
\end{equation}
The fluctuations of $h_{\mu \nu}$  can be expressed as a sum of transverse terms with indices $\;r=0,1\;$, along with some longitudinal, gauge-dependent terms which can only induce  vanishing  fluctuations in the Einstein tensor. In equation,
\begin{align}
\sigma^{in} _{h_{\mu \nu} h_{\rho \sigma}} &\; = \; \sum \sigma^{r~in} _{hh} \pi^{r} _{\mu \nu \rho \sigma}\;+\;\text{longitudinal, gauge-dependent terms}\;, \\
\sigma^{r~in} _{hh} & \;= \;2\pi \kappa \delta (k^2)  \lambda_r\;.
\label{stuff1}
\end{align}
The projectors $\pi^{r} _{\mu \nu \rho \sigma}$   on the transverse space are defined as
\begin{equation}
\pi^r _{\mu \nu \rho \sigma}\; =\; \alpha_r \pi_{\mu \nu} \pi _{\rho \sigma} + \beta_r (\pi_{\mu \rho} \pi_{\nu \sigma}+\pi _{\mu \sigma}\pi_{\nu \rho})\;,
\end{equation}
where
$\;
\pi_{\mu \nu} = \eta_{\mu \nu}-\frac{k_{\mu}k_{\nu}}{k^2}\;
$,
and
$\;\alpha_0 = -\alpha_1 = -\frac{1}{D-1}\;$, $\;\beta_0 = \frac{1}{2}\;$,  $\;\beta_1=0\;$,
$\;\lambda_0 =1\;$, $\;\lambda_1 = -\frac{1}{D-2}\;$.
Being gravitationally induced, the proper fluctuations $\sigma^{in} _{h_{\mu \nu} h_{\rho \sigma}}$ can exist  in the absence of matter.
	
A similar analysis can be carried out for the case of a non-gravitational, vacuum SEM tensor. This tensor has proper fluctuations  even when  $h_{\mu \nu}$ vanishes. As the SEM tensor is  transverse by default, JR decompose it  in terms of the transverse components $\;r=0,1\;$,
\begin{equation}
\sigma _{T_{\mu \nu} T_{\rho \sigma}} ^{in} \;=\; \sum \sigma^{r~in} _{TT} \pi ^{r} _{\mu \nu \rho \sigma}\;,
\label{abovesigmarttin}
\end{equation}
\begin{equation}
\sigma^{r~in} _{TT}\; =\;  \hbar \pi (k^2)^{\frac{D}{2}}  \Theta (k^2)\zeta_r\;,
\label{sigmarttin}
\end{equation}
where, for a massless scalar field,
$\; \zeta_0 = \frac{\Gamma(1+\frac{D}{2})}{(4\pi)^{\frac{D}{2}}\Gamma(D+2)}\;$ and $\;\zeta_1 =\frac{(D-2)^2(D+1)}{2} \zeta_0\;$.

The susceptibility functions for these  proper fluctuations are found to be
\begin{align}
\label{chiA}
\chi^{r~in}_{hh} &\; =\; \frac{2\kappa \lambda_r}{k^2-i\epsilon}\;, \\
\label{chiB}
\chi^{r~in} _{TT} & \;=\;(k^2)^2 \zeta_r \left( \bar{\Gamma}_r + i \pi \hbar (k^2)^{D/2 -2} \Theta(k^2)\right)\;,
\end{align}
where $\;\epsilon\ll k^2\;$ in the former and
 $\bar{\Gamma}_r$ can be left unspecified in the latter,
 as the real part of $\chi^{r~in} _{TT}$ is never needed in what follows.

\subsection{Fluctuations of the stress-energy-momentum tensor}

So far, the  discussion has been limited to only  proper  fluctuations of the metric perturbation and the vacuum SEM tensor. However, there are also ``improper'' fluctuations which can be attributed to the response of these
 same observables to sources.

The linear-response relations for a metric perturbation $h_{\mu \nu}$ responding to a SEM tensor $T_{\mu \nu}$ take the form
\begin{align}
\label{Linear1}
h^r _{\mu \nu} & \;=\; h^{r ~ in} _{\mu \nu} + \chi ^{r~in} _{h h} T^{r} _{\mu \nu}\;,\\
\label{Linear2}
T^r _{\mu \nu} & \;= \;\chi ^{r~in} _{TT} h^{r} _{\mu \nu}\;.
\end{align}
In the latter equation, a term $T^{r~in} _{\mu \nu}$  representing non-gravitational, SEM-tensor fluctuations should have  also been included, but such a term would not influence the  results in the current subsection.

Substituting Eq.~(\ref{Linear1}) into Eq.~(\ref{Linear2}), one obtains
\begin{equation}
T^r _{\mu \nu}\;=\; \frac{\chi ^{r~in} _{TT}}{1-\chi^{r~in}_{hh}\chi^{r~in}_{TT}} h^{r~in} _{\mu \nu}\;.
\label{tchi1}
\end{equation}
The denominator on the right-hand side   can be expanded for low energies $\;\kappa k^2 \ll 1$\;,
leading to
\begin{equation}
T^{r} _{\mu \nu} \;\approx \;\chi_{TT} ^{r~in} h^{r~in} _{\mu \nu}.
\end{equation}

Next, applying the fluctuation-response relation $\text{Im}(\chi ^{r~in}) = \sigma_{TT} ^{r~in}$ to obtain $\;\sigma_{TT} ^{r}\approx\text{Im}(\chi ^{r~in})\;$,   we then have from Eq.~(\ref{sigmarttin}),
\begin{equation}
\sigma_{TT} ^{r}\;=\;\pi \hbar \zeta_r (k^2)^{D/2} \Theta(k^2)\;,
\label{sigamttr1}
\end{equation}
which could then be substituted into  the following analogue   of Eq.~(\ref{abovesigmarttin}):
\begin{equation}
\sigma_{T_{\mu \nu}T_{\rho \sigma}}\; =\; \sum_r \sigma_{TT} ^{r} \pi^r _{\mu \nu \rho \sigma}\;.
\label{sigamttr2}
\end{equation}
	
\subsection{Fluctuations of the metric perturbation}

Our immediate goal  is to formulate the  symmetrized correlation function of the metric perturbation $\sigma_{h_{\mu \nu}h_{\rho \sigma}}$. We  will then use  this correlation function to determine  that of the Einstein tensor $\sigma_{G_{\mu \nu}G_{\rho \sigma}}$, which will tell us about the tensor's fluctuations.

Just like before, we start with the linear-response equations for the metric perturbation,
\begin{align}
h^r _{\mu \nu} &\;=\; \chi ^{r~in} _{hh} T^{r} _{\mu \nu}\;,
\label{htt1}
\\
T^r _{\mu \nu} & \;=\; T^{r~in} _{\mu \nu} + \chi^{r~in} _{TT} h^r _{\mu \nu}\;.
\label{htt2}
\end{align}
However, this time around, it is the proper metric perturbation $h_{\mu\nu}^{r~in}$ that is neglected, as it would not influence the  results in the current subsection.
Similarly, the proper fluctuations of the metric perturbation
do not contribute to fluctuations in the Einstein tensor.

The substitution of Eq.~(\ref{htt2}) into Eq.~(\ref{htt1}) then leads to
\begin{equation}
h^r _{\mu \nu} \;=\; \frac{\chi_{hh} ^{r~in}}{1-\chi_{hh} ^{r~in}\chi_{TT} ^{r~in}} T^r _{\mu \nu}\;\approx\; \chi_{hh} ^{r~in}\left(1+ \chi^{r~in}_{hh}\chi^{r~in} _{TT}\right)T^r _{\mu \nu}\;.
\label{htt3}
\end{equation}
In the second term on the right-hand side, $\;\text{Im}(\chi_{hh} ^{r~in})=2\pi \kappa \lambda_r \delta(k^2)\;$ (see Eqs.~(\ref{sigmachi1}),~(\ref{stuff1})) multiplies $\;\chi^{r~in}_{TT} \propto (k^{D/2})^2\;$ (see Eq.~(\ref{sigamttr1})) to yield  a vanishing contribution. And so
\begin{equation}
h^{r} _{\mu \nu} \;\approx \;\left[\text{Re}(\chi_{hh} ^{r~in})+i\text{Im} (\chi_{hh} ^{r~in})+ \text{Re}(\chi^{r~in}_{hh}) ^2 \chi^{r~in} _{TT}\right] T^r _{\mu \nu}\;.
\label{hmunurr}
\end{equation}

The imaginary part of the square brackets
in Eq.~(\ref{hmunurr}) is, by Eqs.~(\ref{abovesigmachi1}),~(\ref{sigmachi1}),
the symmetrized correlation function $\sigma^{r}_{hh}\;$.
Hence,
\begin{eqnarray}
\sigma ^r _{hh} &=& \text{Im} \left(\text{Re}(\chi_{hh} ^{r~in})+i\text{Im} (\chi_{hh} ^{r~in})+ \text{Re}(\chi^{r~in}_{hh}) ^2 \chi^{r~in}_{TT}\right)
\\ &=& \text{Im} (\chi_{hh}^{r~in}) + \text{Re}(\chi^{r~in}_{hh}) ^2 \text{Im}(\chi^{r~in}_{TT})\;,
\end{eqnarray}
from which it  follows, using the fluctuation-response relation, that
\begin{eqnarray}
\sigma ^r _{hh} &=&  \sigma^{r~in} _{hh} + \text{Re}(\chi^{r~in}_{hh}) ^2 \sigma ^{r~in}_{TT}
\\ &=&
2\pi \kappa \lambda_r \delta(k^2) + \left(\frac{2\kappa \lambda_r}{k^2}\right)^2 \sigma^{r}_{TT}\;,
\end{eqnarray}
where Eqs.~(\ref{stuff1}) and~(\ref{chiA}) have been
used.
Recall that the first term on the right,  the proper fluctuations of the metric perturbation,   does  not make any contribution to
fluctuations in the Einstein  tensor.

\subsection{Relating $\Delta G_{\mu \nu} ^2$ to $\Delta T_{\mu \nu}^2 $}

Still following JR,  one  can relate the Einstein tensor  and the metric perturbation by using the following expressions:
\begin{equation}
\label{EinsteinTensor}
G_{\mu \nu}\; =\; R_{\mu \nu}-\frac{1}{2}\eta_{\mu \nu}R = \eta_{\mu \nu \rho \sigma} R^{\rho \sigma}\;,
\end{equation}
where
\begin{equation}\label{EinsteinTensor2}
\eta_{\mu \nu \rho \sigma}\;=\;\frac{1}{2}\left( \eta_{\mu \rho}\eta_{\nu \sigma} + \eta_{\mu \sigma} \eta_{\nu \rho}-\eta_{\mu \nu} \eta_{\rho \sigma} \right)
\end{equation}
and
\begin{equation}
\label{EinsteinTensor3}
R_{\mu \nu}\;=\;\frac{1}{2}\left( k^2 h_{\mu \nu} + k_{\mu} k_{\nu} \eta^{\alpha \beta} h_{\alpha \beta} - k_{\mu} k^{\sigma} h_{\nu \sigma}-k_{\nu} k^{\sigma} h_{\mu \sigma}  \right)\;.
\end{equation}

After some algebra and recalling that the proper metric fluctuations do not contribute to $\sigma_{G_{\mu \nu} G_{\rho \sigma}}$ , one obtains
\begin{equation}
\sigma_{G_{\mu \nu} G_{\rho \sigma}}\; =\; \hbar \pi \kappa ^2 \left( k^2\right)^{\frac{D}{2}} \Theta(k^2)\sum \zeta_r \pi^{r} _{\mu \nu \rho \sigma}\;,
\end{equation}
which can be identified with the fluctuations of the SEM tensor given in Eqs.~(\ref{sigamttr1}) and~(\ref{sigamttr2}),
\begin{equation}
\sigma _{T_{\mu \nu} T_{\rho \sigma}}\; = \;\hbar \pi (k^2)^{\frac{D}{2}}  \Theta (k^2)\sum \zeta_r \pi ^{r} _{\mu \nu \rho \sigma}\;.
\end{equation}
This yields the desired result for the Einstein-tensor fluctuations,
\begin{equation}
\sigma_{G_{\mu \nu}G_{\rho \sigma}}\; =\; \kappa^2 \sigma_{T_{\mu \nu}T_{\rho \sigma}}\;.
\end{equation}

In terms of correlation functions, the previous identity translates
into (see Eq.~(\ref{correlationFunction1}))
\begin{equation}
C_{G_{\mu \nu}G_{\mu \nu}} (k) \;=\; \kappa^2 C_{T_{\mu \nu}T_{\mu \nu} }(k)\;.
\end{equation}
Keeping in mind that the argument of these functions is really
$|k|$, one  can rewrite the above as
\begin{equation}
C_{G_{\mu \nu}G_{\mu \nu}} (|x|) \;=\; \kappa^2 C_{T_{\mu \nu}T_{\mu \nu} }(|x|)\;.
\end{equation}
Finally,
setting $\;|x|=0\;$, and subtracting the square of  the corresponding Einstein equation, we arrive at
\begin{equation}
		\Delta G_{\mu \nu} ^2 \;=\; \kappa ^2 \Delta T_{\mu \nu }^2\;.
\end{equation}

The normal ordering of operators  does not affect this conclusion  because, as the relevant operators
are standard annihilation and creations operators, the procedure can
modify the zero-point energy but not
 the fluctuations.

\end{document}